\documentclass[sigconf, nonacm]{acmart}
\usepackage{multirow}
\usepackage{multicol}
\usepackage{enumitem}
\usepackage[ruled,vlined]{algorithm2e}
\AtBeginDocument{%
  }

\begin{document}

\title[Clustered Federated Learning with Hierarchical Knowledge Distillation]{Clustered Federated Learning\\ with Hierarchical Knowledge Distillation}

\author{Sabtain Ahmad}
\email{sabtain.ahmad@tuwien.ac.at}
\orcid{0000-0002-1825-0097}
\affiliation{%
  \institution{TU Wien}
  \city{Vienna}
  \country{Austria}
}

\author{Meerzhan Kanatbekova}
\email{meerzhan.kanatbekova@tuwien.ac.at}
\affiliation{%
  \institution{TU Wien}
  \city{Vienna}
  \country{Austria}
}

\author{Ivona Brandi\'{c}}
\email{ivona.brandic@tuwien.ac.at}
\orcid{0000-0002-0668-0035}
\affiliation{%
  \institution{TU Wien}
  \city{Vienna}
  \country{Austria}
}

\author{Atakan Aral}
\email{atakan.aral@univie.ac.at}
\orcid{0000-0001-5318-6422}
\affiliation{%
  \institution{University of Vienna}
  \city{Vienna}
  \country{Austria}
}

\begin{abstract}
  Clustered Federated Learning (CFL) has emerged as a powerful approach for addressing data heterogeneity and ensuring privacy in large distributed IoT environments. By clustering clients and training cluster-specific models, CFL enables personalized models tailored to groups of heterogeneous clients. However, conventional CFL approaches suffer from fragmented learning for training independent global models for each cluster and fail to take advantage of collective cluster insights. This paper advocates a shift to hierarchical CFL, allowing bi-level aggregation to train cluster-specific models at the edge and a unified global model at the cloud. This shift improves training efficiency yet might introduce communication challenges. To this end, we propose CFLHKD, a novel personalization scheme for integrating hierarchical cluster knowledge into CFL. Built upon multi-teacher knowledge distillation, CFLHKD enables inter-cluster knowledge sharing while preserving cluster-specific personalization. CFLHKD adopts a bi-level aggregation to bridge the gap between local and global learning. Extensive evaluations of standard benchmark datasets demonstrate that CFLHKD outperforms representative baselines in cluster-specific and global model accuracy and achieves a performance improvement of 3.32-7.57\%.
\end{abstract}




\maketitle

\section{Introduction}
\label{intro}
Clustered Federated Learning (CFL) provides a powerful framework for collaborative learning in large-scale distributed Internet of Things (IoT) environments. CFL groups clients with similar data distributions together to train cluster-specific global models through device-cloud cooperation while keeping client local data private \cite{ghosh2020efficient, 9174890, ghosh2022efficient, li2021federated, liu2024casa}. This approach is particularly well-suited for IoT applications where geographically distributed clients, though heterogeneous, exhibit inherent clusterability \cite{ouyang2021clusterfl, liu2024casa}. For instance, vehicles operating in similar geographic regions generate spatiotemporally correlated data \cite{shenaj2023learning}. By exploiting such natural clusterability of heterogeneous clients, CFL trains personalized models tailored to specific clusters, making it effective for applications such as autonomous vehicles, smart cities, and healthcare systems \cite{khodak2019adaptive, li2023high, zhou2023hierarchical}. 

\begin{figure}[!t]
  \centering
  \includegraphics[width=\columnwidth]{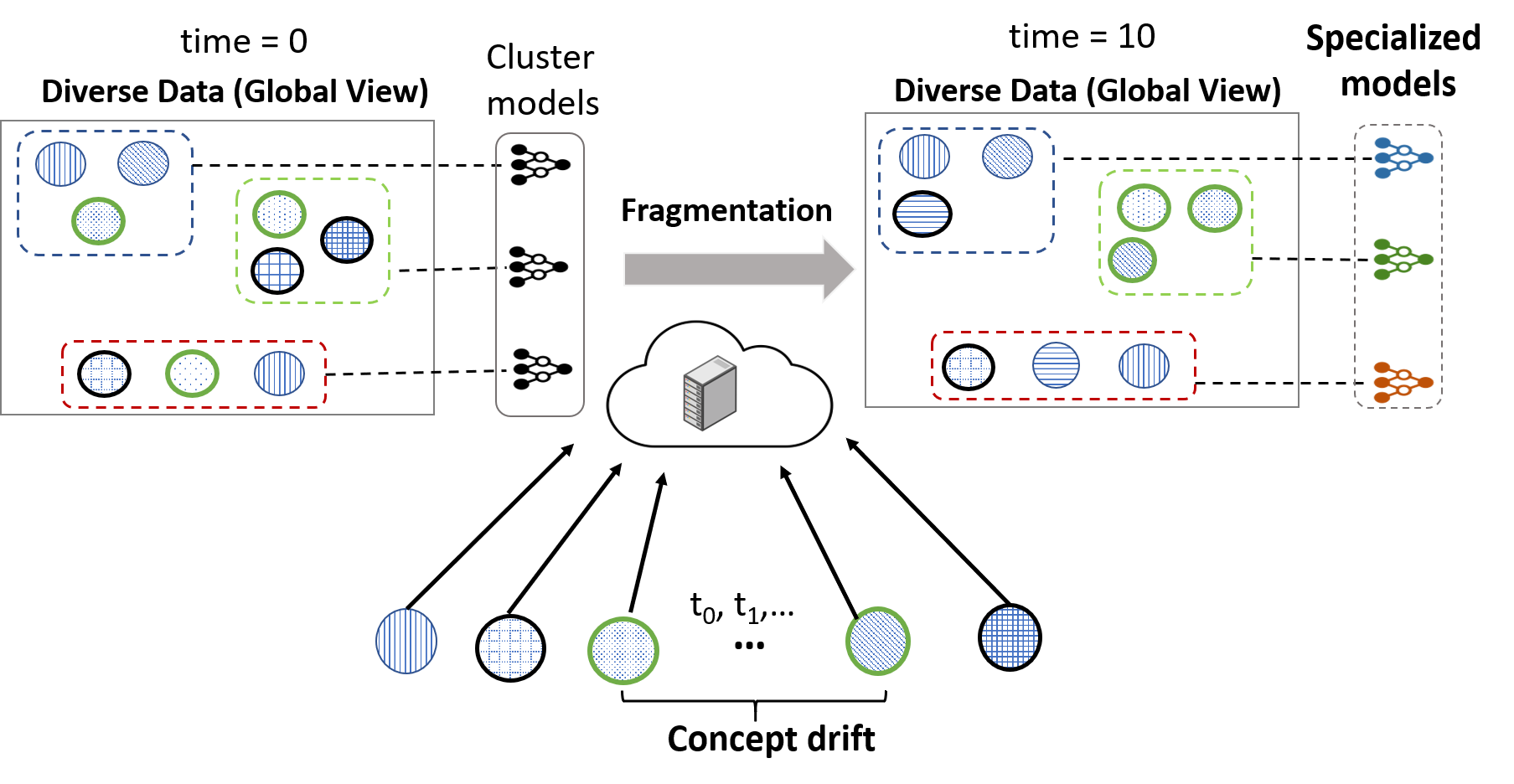}
  \caption{Cluster models with limited data diverge over time and fail to leverage collective shared knowledge.}
  \label{fig:fragmentation_cfl}
\end{figure}

Despite its relative effectiveness in handling data heterogeneity, CFL suffers from several critical limitations that hinder its scalability and adaptability in large-scale IoT environments. Conventional CFL approaches independently train global models for each cluster, with no mechanism for inter-cluster knowledge sharing. As illustrated in Figure \ref{fig:fragmentation_cfl}, these cluster models learn from a subset of clients from a broad representative dataset. However, over time (right side, time = 10), due to concept drift and the absence of cross-cluster knowledge sharing, cluster models become increasingly specialized, leading to fragmented learning and limiting the generalizability of the overall system. 
Moreover, existing CFL methods assess similarity between clients using only model weights. While computationally inexpensive, this approach overlooks the dynamic nature of IoT systems, where client data distributions can shift over time. Consequently, this static similarity metric fails to address concept drift, further exacerbating the challenges of maintaining effective clusters. For example, as clients move or their data distributions evolve, fixed clusters lose relevance, resulting in degraded model performance and reduced adaptability to changing environments.

A promising solution to address these limitations is to adopt a hierarchical or bi-level aggregation framework, which has been used in FL for training a single global model \cite{liu2020client, briggs2020federated, zhou2023hierarchical}. In this framework, cluster-specific models are trained at the edge with device-edge coordination, while a unified global model is trained at the cloud with edge-cloud synergy. This bi-level aggregation ensures both local personalization and global generalization, addressing the lack of global perspective in traditional CFL. Although bi-level aggregation obviates the problem of fragmented learning, simple aggregation may severely impact the effectiveness of the global model. Traditional aggregation strategies such as FedAvg \cite{mcmahan2017communication} struggle in the face of heterogeneity, where clients are geographically and contextually diverse, and may dilute cluster-specific adaptations by averaging model parameters across clusters. Furthermore, overwriting cluster models with the updated global model in every aggregation cycle risks erasing the unique contexts-specific characteristics, ultimately diminishing the benefits of localized training. More specifically, hierarchical CFL faces the following challenges;

\begin{enumerate}
    \item \textit{How to ensure effective CFL with bi-level aggregation?} CFL traditionally manages client clustering and model aggregation at the cloud, converging effectively in static settings. However, bi-level aggregation requires ensuring that the cluster-specific and global models align across different levels without compromising convergence or performance. 
    \item \textit{How to enable knowledge sharing between clusters while preserving cluster-specific characteristics?} Inter-cluster knowledge sharing requires knowledge assimilation from heterogeneous clusters and ensuring that the knowledge distillation does not degrade the personalized performance of individual clusters.
\end{enumerate}

In this paper, we present CFLHKD (Clustered Federated Learning with Hierarchical Knowledge Distillation), a novel methodology for bi-level aggregation in CFL. We introduce novel metrics based on information theory to quantify data heterogeneity across clients, enabling an informed and dynamic response to non-IID data. Based on these metrics, we propose a clustering algorithm that adapts to client mobility and concept drift by grouping clients with similar data distributions to train cluster-specific models and a unified global model while enabling inter-cluster knowledge sharing. We employ FedAvg exclusively for intra-cluster aggregation, as it is better suited for homogeneous settings. To address inter-cluster heterogeneity, we introduce a federated transfer learning (FTL) approach based on multi-teacher knowledge distillation (MTKD), facilitating inter-cluster knowledge sharing while preserving the uniqueness of cluster-specific models. 

\noindent\textbf{\textit{Our contributions:}} Our major contributions are as follows:

\begin{itemize}
\item We propose CFLHKD, one of the first frameworks to address fragmented learning in CFL, enhancing scalability and adaptability in large-distributed environments.  

\item CFLHKD introduces bi-level aggregation leveraging the strengths of FedAvg and MTKD to enable inter-cluster knowledge sharing while preserving personalization.

\item Evaluations on standard benchmarks demonstrate that CFLHKD achieves up to 7.57\% improvement in cluster-specific and global model accuracy compared to baseline methods. 
\end{itemize}


\section{Related Work}
\label{backgound}
\subsection{Clustered Federated Learning} 
CFL addresses one of the fundamental issues of FL, which is the data heterogeneity inherent in most real-world applications, by segregating heterogeneous clients into homogeneous clusters \cite{ghosh2020efficient, 9174890, ghosh2022efficient, li2021federated, liu2024casa}. This clustering-based approach facilitates personalization while maintaining data privacy, making it particularly suitable for heterogeneous IoT environments \cite{he2023clustered, li2021federated, liao2024predicting}. Key areas of CFL research include client similarity metrics and clustering algorithms, both aimed at improving clustering accuracy and its integration with model training. Common client similarity metrics include cosine similarity \cite{duan2021fedgroup}, Euclidean distance \cite{briggs2020federated}, and KL divergence \cite{ma2022convergence}. While on the other hand, clustering methods range from traditional k-means \cite{ghosh2020efficient} to more advanced techniques such as hierarchical and spectral clustering \cite{briggs2020federated, ma2024feduc}. For example, CFL \cite{ghosh2020efficient} iteratively partitions clients based on model weights until training converges. ICFL \cite{briggs2020federated} and FedUC \cite{ma2024feduc} incrementally adapt the clusters using spectral and hierarchical clustering, respectively. However, these methods operate predominantly in single-level aggregation frameworks, where all clustering and aggregation occur at the cloud. As such, they fail to explore the potential of hierarchical aggregation, resulting in fragmented learning with no unified global model.

\subsection{Hierarchical Federated Learning (HFL)}
HFL improves scalability and mitigates non-IID challenges of FL by introducing a hierarchical structure where client models are first aggregated at the edge servers before getting aggregated at the cloud \cite{liu2020client, abad2020hierarchical, liu2022hierarchical,10061835}. For instance, \cite{duan2020self} proposed an HFL framework to address training bias caused by data divergence, while,  \cite{mhaisen2021optimal} compared the impact of data heterogeneity in HFL and centralized FL. Liu et al. \cite{liu2020client} introduced a framework that focuses on tackling client heterogeneity by leveraging multi-level aggregation. 
Existing HFL frameworks have primarily focused on training a single global model, often assuming that clients are statically grouped and assigned to specific edge servers \cite{liu2020client, liu2022hierarchical, 10459855}. Although this approach has proven effective for addressing non-IID data and improving scalability, these methods have not been adapted for integration with CFL. Consequently, they fail to address the fragmented learning problem inherent to CFL, where cluster-specific models are trained independently without inter-cluster knowledge sharing.

Furthermore, the challenge of aggregating heterogeneous cluster-specific models at the cloud while preserving their unique characteristics remains unexplored. In bi-level aggregation, edge servers generate cluster-specific models, while the cloud aggregates these models to form a unified global model. This transition demands solutions to address two key issues: 1) effective aggregation of heterogeneous cluster models, where standard methods like FedAvg fail due to inter-cluster heterogeneity, and 2) ensuring consistency between edge-level clustering and cloud-level model generation. 

Our work addresses these challenges by introducing CFLHKD, a hierarchical CFL framework. Using multi-teacher knowledge distillation (MTKD), CFLHKD enables effective inter-cluster knowledge sharing without overriding cluster-specific characteristics. Additionally, we propose a dynamic aggregation strategy at the cloud that accounts for the heterogeneity of cluster-specific models, ensuring the generation of a robust unified global model. Together, these innovations bridge the gap between localized cluster learning and global model generalization, overcoming the limitations of conventional CFL and HFL approaches.

\section{Problem Statement}
\label{problem}

This section formalizes the challenges of conventional CFL and motivates the transition to a hierarchical CFL (H-CFL). We describe the H-CFL workflow and outline its potential to bridge inter-cluster gaps while preserving cluster specialization. Subsequently, we define the scope of our work, highlighting key constraints and considerations for deploying H-CFL in real-world scenarios.

\begin{figure}[!t]
  \centering
  \includegraphics[width=\columnwidth]{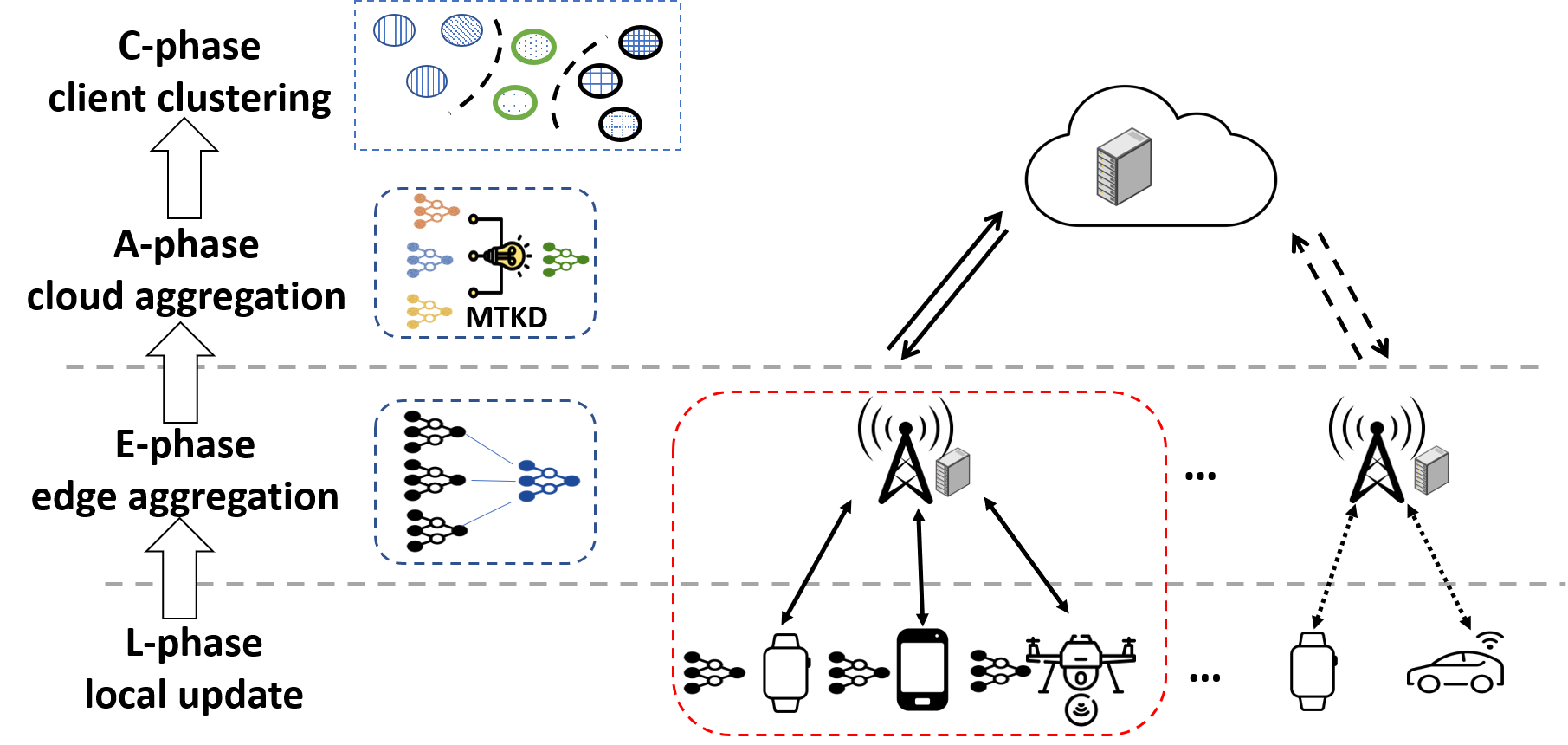}
  \caption{H-CFL workflow.}
  \label{fig:workflow}
\end{figure}

\textbf{H-CFL Workflow:} 
In H-CFL, clients collaboratively learn both cluster-specific models and a unified global model through a hierarchical, bi-level optimization, iteratively minimizing the clustering error and learning objectives until model convergence. The H-CFL workflow operates across multiple levels; clients, edge servers, and the cloud, and involves four steps (see Figure \ref{fig:workflow}): i) Local Training (L-phase), in which each client performs local model training and sends the local updates to its assigned edge server; ii) Edge Aggregation (E-phase), where edge servers aggregate client updates to generate cluster-specific models for each cluster followed by sending the local updates or cluster-specific models to the cloud; iii) Cloud Aggregation (A-phase), the central cloud aggregates the cluster-specific models and generates a unified global model; iv) Client Clustering (C-phase), wherein the cloud adjusts and groups clients based into clusters based on local model updates.

Formally, let $n$ clients $\{c_1, \cdots, c_n$\} with local datasets $\{D_1,  \cdots, D_n\}$ be grouped into $K$ clusters $\{u_1, \cdots, u_k$\}. All clients with a cluster $u_k$ collaboratively train a cluster-specific model $w_{e,k}$, while the cloud server trains a unified global model $w_g$. The H-CFL training process minimizes the following objectives:
\begin{equation}\label{eq:edge_obj}
    \min_{w_{e,1}, \cdots, w_{e,k}} \mathcal{P}_{edge} = \sum_{k=1}^K \frac{1}{|C_k|}\sum_{c_i \in C_k} E [\mathcal{L}(w_{e,k}; D_i)]
\end{equation}

where $C_k$ denotes the set of clients in cluster $u_k$. Similarly, the goal of training a unified global model is translated to minimizing the following objective:
\begin{equation}\label{eq:global_obj}
    \min_{w_{g}} \mathcal{P}_{cloud} = \frac{1}{n}\sum_{i=1}^n E [\mathcal{L}(w_{g}; D_i)]
\end{equation}

where $w_g$ denotes the unified global model. Meanwhile, H-CFL also minimizes the following clustering error:
\begin{equation}\label{eq:cluster_obj}
    \min_{w_{1}, \cdots, w_{n}, w_{e,1},\cdots, w_{e,k}} \mathcal{H} = \sum_{k=1}^K \frac{1}{|C_k|}\sum_{c_i \in C_k} ||w_i - w_{e,k} ||_2^2
\end{equation}

where $w_i$ and $w_e,k$ represent the local model of client $c_i$ and the cluster-specific model, respectively.

\begin{figure}[!t]
  \centering
  \includegraphics[width=\columnwidth]{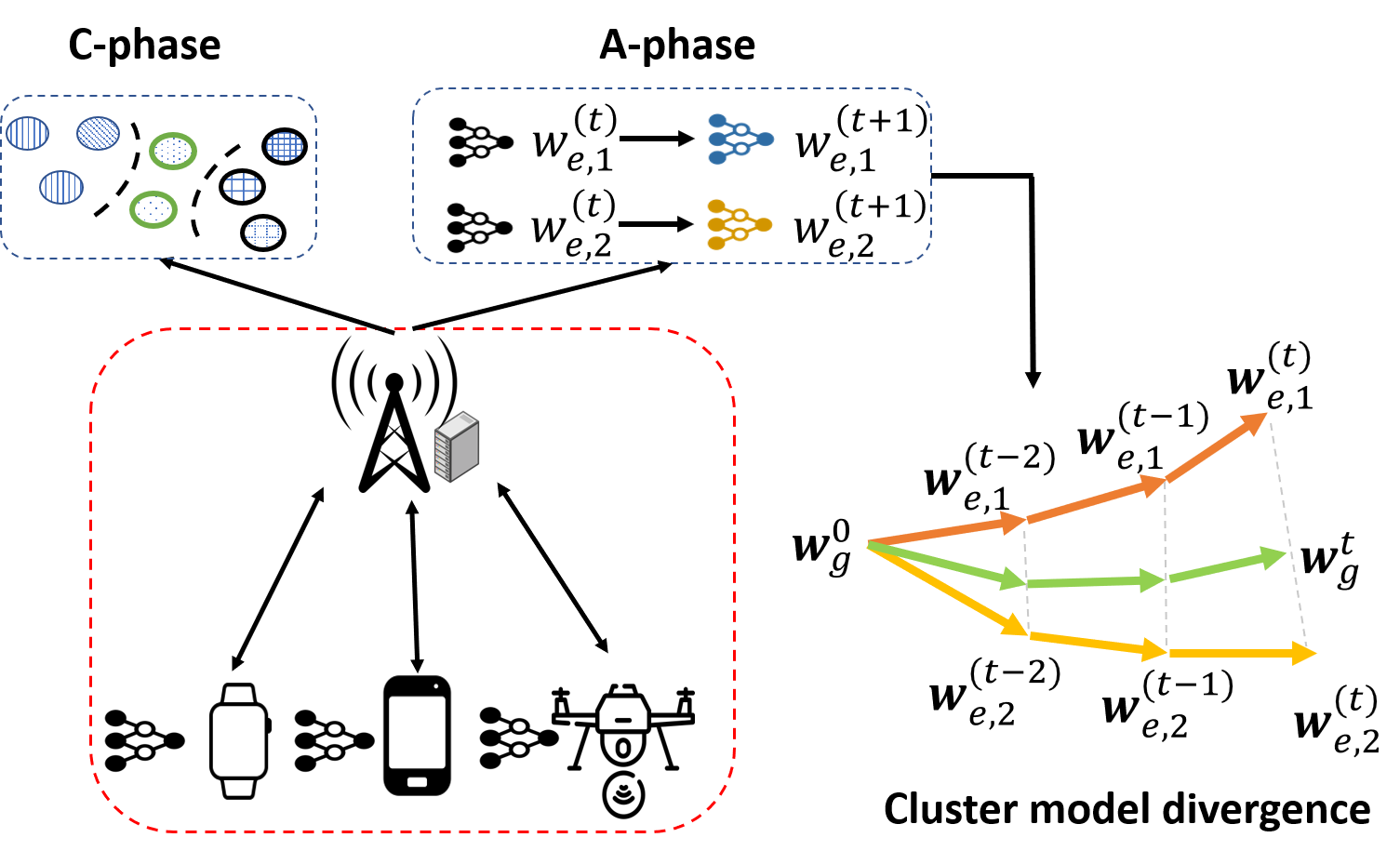}
  \caption{CFL: Cluster model divergence and challenge to generate a global model.}
  \label{fig:cfl_challenge}
\end{figure}

\textbf{Transition to H-CFL:} As mentioned in Section \ref{intro}, we advocate a transition from CFL to H-CFL to overcome the inherent limitations of fragmented learning in CFL (highlighted in Figure \ref{fig:cfl_challenge}). The inability of CFL to produce a unified global model and facilitate inter-cluster knowledge sharing calls for substantial modifications to both the aggregation and cluster model update.

\begin{itemize}
    \item \textit{Model Aggregation:} Model aggregation in CFL is limited to cluster level, where each cluster trains its own model independently \cite{ghosh2020efficient, 9174890}. This lack of inter-cluster knowledge sharing coupled with the absence of a unified global model hinders generalization, particularly in scenarios where certain clusters suffer from insufficient data diversity. Aggregating high-level shared knowledge is essential to address this limitation, but its effectiveness depends on balancing cluster personalization and global generalization.
    \item \textit{Model Update:} Conventional CFL methods \cite{ghosh2022efficient, li2021federated, liu2024casa} overwrite cluster-specific models at each iteration by aggregating local client updates, discarding the previous model entirely. This approach fails to leverage historical knowledge and shared insights, necessitating a new strategy to retain cluster-specific characteristics while integrating global knowledge through controlled knowledge distillation.
\end{itemize}

\textbf{Assumptions and Scope:} We assume that the same number of edge servers as the number of clusters are strategically placed to ensure low-latency communication with their assigned clients. While this placement is assumed to be predetermined, additional methods, such as multi-objective optimization, could be explored to dynamically optimize edge server locations for improved performance \cite{sabtainIoT23}. Our design emphasizes server-side cloud and edge modifications to maintain practical deployment, minimizing changes to client workflows. FedAvg \cite{mcmahan2017communication} is used at the edge due to the homogeneity of client clusters, while MTKD is employed at the cloud to balance global generalization and local specialization. Although privacy remains integral to FL; this work focuses on improving training efficiency, scalability, and accuracy. Specific explorations of privacy attacks or defenses are beyond the scope of this work.

\section{Method}
\label{fl_problem}
This section presents our proposed CFLHKD (Clustered Federated Learning with Hierarchical Knowledge Distillation), a hierarchical CFL scheme. We begin by analyzing the limitations of conventional CFL in producing a unified global model and enabling inter-cluster knowledge sharing (Sec. \ref{limitations_cfl}). We then introduce our bi-level aggregation strategy, with cluster-specific models trained at the edge and global knowledge integrated at the cloud through MTKD (Sec. \ref{bi-level}). To preserve cluster-specific characteristics while incorporating global knowledge, we propose a federated transfer learning (FTL)-based distillation mechanism (Sec. \ref{ftl}). 

\subsection{Understanding Fragmented Learning in CFL}
\label{limitations_cfl}
Although CFL partially mitigates the non-IID challenge by grouping clients into clusters, it inherently suffers from fragmented learning; isolated cluster models lack cross-cluster coordination. As illustrated in Figure \ref{fig:cfl_challenge}, this isolation prevents inter-cluster knowledge sharing and results in the absence of a unified global model.

Empirical evidence (see Sec. \ref{results}) shows that training in isolated clusters degrades model performance when tested on broader, unseen distributions, with accuracy drops of up to 6.8\% compared to models trained under inter-cluster collaboration. This section formalizes how fragmented learning emerges and its implications for CFL optimization.

\textbf{Direct Impact:}  While CFL ensures intra-cluster similarity, it overlooks potential relationships between clusters, exacerbating inter-cluster disparity. Without a mechanism to align cluster-specific models with global objectives, clusters diverge. Specifically, let two clusters $u_1, u_2$ with client sets $C_1, C_2$ and distinct data distributions $D_1, D_2$. Each cluster independently trains its model:
\begin{equation}
    w_{e,k}^{(t)} = w_{e,k}^{t-1} - \eta \nabla \mathcal{L}_k(t-1)
\end{equation}

where $\mathcal{L}_k$ is $k-th$ cluster's training loss. Without cross-cluster coordination, cluster models gradually diverge over time, with the distance between models bounded by KL divergence of their underlying data distributions:
\begin{equation} \label{eq:model_divergence}
    ||w_{e,1}^{(t)} - w_{e,2}^{(t)}||^2 \propto \Delta_D
\end{equation}

where $\Delta_D = KL(D_1||D_2) + KL(D_2||D_1)$. As $\Delta_D$ increases, clusters learn increasingly distinct features. Without cross-cluster coordination, models amplify distributional differences, which impacts generalization on tasks requiring shared knowledge. 

Furthermore, clusters with overlapping features (e.g., shared classes in image classification) independently learn redundant representations. For $K$ clusters with $S$ overlapping features, this redundancy increases computation and communication costs by $\mathcal{O}(S.K)$.

\textbf{Compound Impact: } The interplay between these challenges compromises CFL's ability to jointly optimize the training objectives (Equations \ref{eq:edge_obj} and \ref{eq:global_obj}). The lack of a unified global model reduces the effectiveness of learning across clusters, particularly when cluster sizes are imbalanced or clusters have overlapping distributions. Specifically, the global loss $\mathcal{P}$ fluctuates as the clusters overfit to local data. Let $\mathcal{P}_k$ denote the loss of cluster $u_k$. The variance of the global loss grows as:
\begin{equation} \label{obj_instabaility}
    Var(\mathcal{P}) = \sum_{k=1}^K p_k (\mathcal{P}_k - E[\mathcal{P}])^2
\end{equation}

where $p_k$ is the weight of cluster $u_k$. High variance slows convergence and degrades final accuracy.

Moreover, misaligned cluster models exacerbate clustering errors and lead to a feedback loop. For instance, misaligned cluster models result in poor cluster assignments, leading to higher clustering error $\mathcal{H}$ (Equation \ref{eq:cluster_obj}), which in turn leads to worsening of the divergence of the models. Let $r^t$ denote the mis-clustering rate at round $t$. The resulting feedback loop increases $r$ over time:
\begin{equation} \label{eq:clustering_error}
    r^{t+1} \geq r^t + \frac{\eta}{K} \sum_{k=1}^K ||w_{e,k}^t - w^*_g||^2 
\end{equation}

where $w^*_g$ is the optimal global model. This creates a vicious cycle where clustering errors increase model divergence.

\subsection{Bi-level Hierarchical Aggregation}
\label{bi-level}
In large distributed IoT environments, it is crucial to ensure efficient and scalable training of predictive models for enabling real-time IoT applications. However, FL faces significant challenges when applied to these ecosystems due to inherent data heterogeneity across devices and regions, stemming from non-IID data patterns that arise naturally in most IoT applications. Although CFL addresses non-IID by training cluster-specific individual models, it suffers from fragmented learning. To address this, we propose bi-level hierarchical aggregation framework where edge servers resolve intra-cluster heterogeneity before cluster-cloud coordination. This bi-level hierarchical aggregation leverages the strengths of FedAvg to generate task-specific cluster models at the edge using the relatively homogeneous set of clients and multi-teacher knowledge distillation (MTKD) for assimilating the diverse knowledge from heterogeneous cluster models to enable inter-cluster knowledge sharing. More specifically, each client $c_i$ in cluster $u_k$ trains a local model $w_i$ on its private dataset $D_i$, minimizing local loss (L-phase):
\begin{equation}\label{eq:local_obj}
    \min_{w_i} \mathcal{L}_{local}(w_i, D_i)_=  \frac{1}{|D_i|} \sum_{(x,y) \in D_i} {\ell}(f(x;w_i), y))
\end{equation}

where $f(x;w_i), y)$ is the $i-th$ client model's output for input $x$.

In H-CFL, the first level of aggregation between clients and the corresponding edge server typically operates within well-defined and localized environments. In such settings, data from clients under a single edge server is relatively homogeneous; therefore, traditional parameter-based aggregation methods, such as FedAvg \cite{mcmahan2017communication}, work well. The edge server for cluster $u_k$ aggregates local models into a cluster-specific model $w_{e,k}$, weighted by client data sizes to account for intra-cluster heterogeneity:
\begin{equation} \label{eq:edge_agg}
    w_{e,k} = \sum_{c_i \in C_k} \frac{|D_i|}{|D_k|} w_i, \quad \text{where } |D_k| = \sum_{c_i \in C_k} |D_i|
\end{equation}

This edge-aggregation (E-phase) minimizes the cluster-level empirical risk; effectively solving a federated optimization problem within $C_k$:
\begin{equation} \label{eq:e-phase}
    \mathcal{P}_{edge}(w_{e,k}) = \sum_{c_i \in C_k} \frac{|D_i|}{D_k} \mathcal{L}_{local}(w_i, D_i)
\end{equation}

In conventional CFL, each cluster $u_k$ trains a cluster-specific model via client updates directly at the cloud. This single-level aggregation leads to 1) model divergence across clusters and 2) lack of inter-cluster coordination, reinforcing fragmented learning. H-CFL replaces single-level aggregation with bi-level hierarchical aggregation, edge aggregation (handles intra-cluster heterogeneity), and cloud aggregation (fuses inter-cluster knowledge).

\textbf{Cloud-Aggregation:} At the cloud level, where task-specific models from different regions are combined, the heterogeneity of these models reflecting diverse data distributions and task characteristics due to different geographical and contextual conditions renders vanilla FedAvg essentially ineffective. Traditional FL methods (e.g., FedAvg) aggregate cluster models as:
\begin{equation}
    w_g = \sum_{k=1}^K \frac{|D_k|}{D} w_{e,k}
\end{equation}

However, this approach assumes that all clusters share the same data distributions and that each cluster contributes equally, ignoring data size/quality, model divergence, and relevance to the global model. These assumptions break down in CFL because clusters specialize in localized distributions.

Building on edge-aggregation cluster specialization, the cloud synthesizes cross-cluster knowledge into a unified global model $w_g$, mitigating fragmented learning while balancing contributions from heterogeneous clusters. The cloud aggregates cluster-specific models $\{w_{e_,k}\}_{k=1}^K$ into $w_g$ by applying a dynamically weighted aggregation that accounts for  data richness ($|D_k|$), cluster model accuracy ($\alpha_k$), and divergence from the global objective: 
\begin{equation} \label{eq:a-phase}
    w_g = \sum_{k=1}^K \rho_k w_{e,k} 
\end{equation}

To address imbalanced data quality or task relevance, weights $\rho_k$ are refined using cluster performance metrics (e.g., validation accuracy $\alpha_k$):
\begin{equation} \label{eq:aggregation}
    \rho_k = \frac{|D_k|.\alpha_k e^{-\lambda ||w_{e,k} - w_g||^2}}{\sum_{j=1}^K |D_j|.\alpha_j e^{-\lambda ||w_{e,j} - w_g||^2}}.
\end{equation}

This ensures that clusters with better generalization influence the global model more while clusters too distant from $w_g$ are penalized. Empirical results (see Sec. \ref{results}) demonstrate that the dynamic weighting of the H-CFL improves the accuracy of the global model by 5\% compared to simple averaging, while reducing communication overhead by 12\%.

\begin{algorithm}[t]
\caption{CFL via hierarchical knowledge distillation}
\label{alg:FDC}
\SetAlgoLined
\DontPrintSemicolon
\KwIn{Rounds $T$, clients \( C = \{ c_1, c_2, \dots, c_n \} \)}
\KwOut{Global model $w_g$, cluster models $\{w_{e,k}\}_{k=1}^K$}

Initialize global model $w_g^{(0)}$ and cluster assignments $\{C_k\}_{k=1}^K$\;

\For{each round $t = 1$ to $T$}{
    \tcc{1. Local Training (L-phase)}
    Each client $c_i \in C_k$ optimizes its model using Eq. \ref{eq:local_obj}.

    \tcc{2. Edge Aggregation (E-phase)}
    Each cluster $k$ aggregates client models using Eq. \ref{eq:e-phase}.

    \tcc{3. Cloud Aggregation (A-phase)}
    Compute dynamic weights $\rho_k$ and update the global model using Eq. \ref{eq:a-phase} and Eq. \ref{eq:aggregation}.

    \tcc{4. Cluster Refinement}
    Each cluster is refined via global model using Eq. \ref{eq:cluster_refine}.

    \tcc{5. Dynamic Clustering (every $T_{\text{cluster}}$ rounds)}
    \If{$t \bmod T_{\text{cluster}} = 0$ or $JSD(Q_{c_n}^t || Q_{c_n}^{t+\Delta t})> \eta$ (drift)}{
        Compute affinity matrix $\mathbf{A}$ using Eq.\ref{eq:affinity}, sort clients, and reassign them to clusters based on affinity.
    }
}
\end{algorithm}

\subsection{Global-Guided Cluster Refinement}
\label{ftl}
To harmonize cluster specialization with cross-cluster generalization, we introduce a global-guided refinement phase, where edge models $w_{e,k}$ are fine-tuned using the global model $w_g$. This step mitigates fragmented learning by incorporating shared global knowledge while preserving cluster-specific characteristics. 


Cluster-specific models $w_{e,k}$ are refined via a regularized loss function that balances local performance and global alignment:
\begin{equation} \label{eq:ftl}
    \min_{w_{e,k}} P_{\text{FTL}} = \underbrace{\mathbb{E}\left[\mathcal{L}(w_{e,k}; \mathcal{D}_k)\right]}_{\text{Cluster Specialization}} + \underbrace{\lambda ||w_{e,k} - w_g||^2}_{\text{Global Generalization}}
\end{equation}

where $\lambda>0$ is a hyperparameter that controls the influence of the global model $w_g$. Cluster-specific models are fine-tuned using gradient descent: 
\begin{equation} \label{eq:cluster_refine}
    w_{e,k}^{(t+1)} = w_{e,k}^{(t)} - \eta_k \nabla \left[ \mathcal{L}(w_{e,k}^{(t)}; \mathcal{D}_k) + \lambda ||w_{e,k}^{(t)} - w_g||^2 \right],
\end{equation}

We refine $\lambda$ per cluster using a divergence-aware weighting:
\begin{equation}
    \lambda_k = \frac{\lambda_0}{1+div(w_{e_k}, w_g)}
\end{equation}

where $div$ measures cosine distance between $w_{e,k}$ and $w_g$

Unlike conventional CFL’s hard aggregation (overwriting $w_{e,k}$ with $w_g$), FTL enables inter-cluster knowledge sharing while preserving cluster-specific features through regularization.

\subsection{Federated Dynamic Clustering}

Traditional CFL assumes static client assignments, leading to suboptimal cluster formation in dynamic environments. Clients with similar data distributions may be assigned to different clusters due to evolving data distributions, while heterogeneous clients within the same cluster degrade model performance. To address this, we propose Federated Dynamic Clustering (FDC), a cloud-tier mechanism to dynamically group clients into edge clusters based on spatio-temporal data distributions and model affinity. FDC ensures cohesive clusters despite concept drift, aligning with the H-CFL goals of mitigating fragmented learning and scalability challenges.


We define a hybrid affinity score that jointly captures clients' data distribution similarity via Jensen-Shanon Divergence (JSD) and model alignments via cosine similarity. For clients $c_i, c_j$, the affinity is computed as:
\begin{equation} \label{eq:affinity}
    \mathcal{A}(c_i, c_j) = \underbrace{\gamma \cdot JSD(Q_{i}||Q_{j})}_{\text{Data Distributions}} + (1 - \gamma) \cdot \underbrace{ \frac{\langle w_{i}, w_{j} \rangle}{\|w_{i}\|_2 \cdot \|w_{j}\|_2}}_{\text{Model Affinity}}
\end{equation}




where JSD quantifies divergence between client data distributions, computed using local histograms (e.g., label frequencies). Model Affinity measures alignment between local models, and $\gamma$ is a hyperparameter controlling the tradeoff (default: $\gamma = 0.5$).

\textbf{Client Clustering (C-phase):} FDC employs a sorted threshold-based clustering algorithm that balances accuracy and efficiency. FDC begins by computing the affinity matrix $\mathbf{A} \in \mathbb{R}^{n \times n}$. Clients are ranked based on their total affinity scores computed as follows:
\begin{equation}
    ||A_i||_2 = \sqrt{\sum_j A(c_i, c_j)^2}
\end{equation}

Clients with higher affinity norms act as stable anchors for cluster formation. Clusters are formed following a threshold-based assignment strategy to ensure high intra-cluster coherence. The first cluster is initialized with the highest ranked client $c_1$. Each subsequent client $c_i$ is assigned to an existing cluster $k$ if its affinity distance to the cluster centroid $\mu_k$ satisfies $dist(c_i, \mu_k) \leq \delta$, where $dist(c_i, \mu_k) = ||A(c_i), \mu_k||$  and $\mu_k = \frac{1}{C_k} \sum_{c_j \in C_k}c_j$. Otherwise, a new cluster is created.

To maintain robust clustering, within-cluster variance is monitored and refined. Specifically, enforce the constraint; $Var_k =\frac{1}{C_k}\sum_{c_i \in C_k} dist(\mu_k, c_i)^2 \leq \delta^2$. The violation of this constraint indicates high intra-cluster divergence, enforcing adaptive cluster merging or splitting. This dynamic adaptation ensures that client assignments remain efficient and stable, preserving the integrity of hierarchical FL across evolving client distributions.

For $k$ clusters, each with ($|C_k| = m$) clients, the \textit{worst-case within sum of squares (WCSS)} is bounded by:
\begin{equation*}
    WCSS \leq \delta^2 (n - m)
\end{equation*}

Each client satisfies $dist(c_i, \mu_k) \leq \delta$; summing over all clusters:
\begin{equation}
\begin{split}
    WCSS & = \sum_{k=1}^k \sum_{c_i \in C_k} (\lVert c_i - c_{1,k} \lVert)^2 \\
    &= \sum_{k=1}^k \sum_{c_i \in C_k, c_i\neq c_{1,k}} (\lVert c_i - c_{1,k} \lVert)^2\\
         & \leq \sum_{k=1}^k (|C_k|-1)\delta^2 = \delta^2(\sum_{k=1}^k |C_k|-m) = \delta^2 (n-m)
\end{split}
\end{equation}

This bound corresponds to the worst-case assignment. In practice, the computed $WCSS$ is at least two times smaller than the worst-case bound \cite{elsworth2020abba}, that is; 
\begin{equation}
    WCSS  = \frac{\delta^2 (n-m)}{2}
\end{equation}

Homogeneous clusters ($WCSS \propto \delta^2$) with low WCSS reduce the gradient variance in the global objective $\mathcal{P}_{cloud}$, accelerating the convergence of H-CFL.

\textbf{Dynamic Adaptation:} During model training, client assignments are monitored for concept drift or mobility. A drift in client's data distribution, detected as $JSD(Q_{c_n}^t || Q_{c_n}^{t+\Delta t})> \phi$; where $\phi$ is drift threshold (typical range set via grid search: $\phi \in [0.1, 0.9]$), triggers re-evaluation of its cluster assignment, as detailed in algorithm \ref{alg:FDC}. Upon reassignment, clients initialize with their new cluster model $w_{e,k'}$ to ensure stability.

\begin{table*}[h]
    \centering
    \renewcommand{\arraystretch}{1.0} 
    \setlength{\tabcolsep}{5pt} 
    \caption{Overall performance: "-" indicates either 0 (in the case of Comm) or failure to reach the target accuracy. 
    }
    \begin{tabular}{l|ccc|ccc|ccc|ccc|ccc}
        \toprule
        \multirow{2}{*}{{Method}} & \multicolumn{3}{c|}{{MNIST}} & \multicolumn{3}{c|}{{CIFAR-10}} & \multicolumn{3}{c|}{{FEMNIST}} & \multicolumn{3}{c|}{{HAM10000}} & \multicolumn{3}{c}{{CITYSCAPES}} \\
        \cline{2-16}
        & {Acc} & Comm &{Time} & {Acc} & Comm & {Time} & {Acc} & Comm & {Time} & {Acc} & Comm & {Time} & {mIoU} & Comm & {Time} \\
        \midrule
        Standalone    & 92.3 & -  & 1.8  & 68.1 & - & - & 84.7 & -  & 3.6  & 62.5 & -  & - & 61.3 & - & - \\ 
        \hline
        FedAvg    & 95.1 & 48.0 & 3.2  & 76.8 & 448.0 & 8.5 & 89.2 & 148.0 & 5.1  & 67.3 & 280 & 6.8 & 66.5 & 448.0 & 22.1 \\
        FedProx   & 95.7 & 43.2 & 3.1  & 77.4 & 403.2 & 8.3 & 89.8 & 133.2 & 5.0  & 69.1 & 252.0 & 6.7 & 68.4 & 403.2  & 21.8 \\
        HierFAVG  & 95.9 & 28.8 & 2.9  & 78.2  & 268.8 & 7.9 & 90.1  & 88.8 & 4.8  & 69.8 & 168.0 & 6.4 & 67.1 & 268.8 & 19.7\\
        FL+HC     & 96.2 & 24.0 & 2.7  & 78.9 & 224.0 & 7.5 & 91.3 & 74.0  & 4.5  & 70.5  & 140.0 & 6.1 & 70.2 & 224.0 & 18.9\\
        CFL       & 96.5 & 19.2 & 2.5  & 79.3 & 179.2 & 7.2 & 91.7 & 59.2 & 4.3  & 71.2 & 112.0 & 5.8 & 70.2 & 179.2 & 17.2\\
        ICFL      & 96.8 & 14.4 & 2.4  & 80.1 &  134.4 & 7.0  & 92.0 & 44.4 & 4.1  & 72.1 & 84.0  & 5.6 & 71.6  & 134.4 & 16.5\\
        IFCA      & 97.0 & 12.0 & 2.3  & 80.5 & 112.0 & 6.8  & 92.4 & 37.0 & 3.9  & 72.8 & 70.0 & 5.4 & 72.1 & 112.0 & 15.8\\
        CFLHKD   & \textbf{97.6} & \textbf{9.6} & \textbf{2.1} & \textbf{82.3} & \textbf{89.6} & \textbf{6.3} & \textbf{93.5} & \textbf{26.6} & \textbf{3.5} & \textbf{78.9} & \textbf{56.0}& \textbf{5.0} & \textbf{ 74.8} & \textbf{89.6}  & \textbf{14.4}\\
        \bottomrule
    \end{tabular}
    \label{tab:overall}
\end{table*}

\begin{table}[h]
    \centering
        \renewcommand{\arraystretch}{1} 
    \setlength{\tabcolsep}{5pt} 
    \caption{Dynamic Non-IID (Concept Drift)}
    \resizebox{\columnwidth}{!}{%
    \begin{tabular}{l|cc|cc|cc}
        \toprule
        \multirow{2}{*}{Method} & \multicolumn{2}{c|}{MNIST} & \multicolumn{2}{c|}{CIFAR-10} & \multicolumn{2}{c}{HAM10000}   \\
        \cline{2-7}
        & Acc Drop & Recovery & Acc Drop & Recovery & Acc Drop & Recovery  \\
        \midrule
        FedAvg    & 12.5 &	15 &	18.3	& -	& 15.8 &	20 \\
        HierFAVG  & 10.5 &	10 &	15.6	& 15	& 13.2 &	14 \\
        IFCA      & 8.2 &	6 &	12.7 &	8 &	10.5 &	7 \\
        CFLHKD   & \textbf{2.1} & \textbf{2} & \textbf{4.8} & \textbf{3} & \textbf{3.2} & \textbf{2} \\
        \bottomrule
    \end{tabular}
    }
    \label{tab:dynamic}
\end{table}



\section{Experiments}
\label{results}
\subsection{Experimental Setup}
We compare our proposed CFLHKD framework against state-of-the-art FL, CFL, and hierarchical FL baselines.  

\begin{itemize}
    \item \textbf{Standalone:} Each client trains its model on its local dataset without any federated coordination.
    \item \textbf{FedAVg\cite{mcmahan2017communication}:} Vanilla FL with global model aggregation.
    \item \textbf{FedProx\cite{li2020federated}:} FL with a proximal term to improve convergence under data heterogeneity.
    \item \textbf{HierFAVG\cite{liu2020client}:} Hierarchical FL with edge-cloud aggregation 
    \item \textbf{FL+HC\cite{briggs2020federated}:} a framework that integrates hierarchical clustering with FL to address non-IID problem.
    \item \textbf{CFL\cite{ghosh2020efficient}:} CFL method that uses gradient-based bi-partitioning. 
    \item \textbf{ICFL\cite{yan2023clustered}:} CFL method that  employs incremental clustering.
    \item \textbf{IFCA\cite{ghosh2022efficient}:} CFL method that iteratively clusters clients based on loss minimization. 
\end{itemize}

\textbf{Dataset:} We evaluate on two tasks under static and dynamic (non-IID) settings. 1) image classification (IC): MNIST \cite{deng2012mnist}, CIFAR-10 \cite{krizhevsky2009learning}, FEMNIST \cite{caldas2018leaf}; partitioned across clients using Dirichlet distribution $(\alpha=0.5)$. HAM10000\cite{tschandl2018ham10000}; skin lesion dataset partitioned by hospital (5 clusters, 20 clients/cluster). 2) semantic segmentation (seg); CityScapes \cite{cordts2016cityscapes} street scenes partitioned by geographical regions (50 cities). Each edge server corresponds to a city, with fleets of vehicles representing clients to simulate real-world non-IID data. 

\textbf{Metrics:} Our evaluation considers three primary metrics: 1) \textit{Accuracy}: test accuracy or mIoU of cluster-specific and global models; 2) \textit{Communication Cost}: total data transferred between edge and cloud; 3) \textit{Time to Convergence}: total training time until convergence.  

Additional experimental details, including hyperparameter settings, are provided in Appendix \ref{add_exp_settings}.

\subsection{Main Results}
\label{main_results}
\subsubsection{Non-IID Settings}

Table \ref{tab:overall} presents the overall performance of CFLHKD alongside the baselines, evaluating them on the final accuracy/mIoU, communication cost (Comm), and convergence time (Time). We report the following key findings.
\begin{itemize}
    \item \textit{Improved performance through inter-cluster knowledge sharing;} CFLHKD consistently achieves the highest accuracy/mIoU across all datasets, outperforming the standard FL, HFL, and CFL baselines. For example, CFLHKD reaches an accuracy of $97.6\%$ on MNIST, a $2.6\%$ improvement over FedAvg, a $0.6\%$ over IFCA, and a $5.3\%$ over Standalone baseline. Similar trends are observed on CIFAR-10 and FEMNIST datasets, achieving an accuracy of $82.3\%$ ($1.8\%$ over IFCA) and $93.5\%$ ($1.2\%$ over IFCA). The advantages of inter-cluster knowledge sharing are particularly pronounced in complex tasks, for instance, on HAM10000, achieved $78.9\%$; $8.4\%$ higher than IFCA, despite significant heterogeneity in lesion imaging across institutes. Similarly, in CITYSCAPES urban segmentation task, CFLHKD achieves $74.8\%$ mIoU, exceeding the 2nd best performing baseline by $3.8\%$.

    \item \textit{Communication efficiency via bi-level aggregation;} CFLHKD reduces communication costs by $20-85\%$ in comparison with baselines by selectively synchronizing cluster-level and global updates. On CIFAR-1O, CFLHKD reduces communication by $20\%$ compared to IFCA. For tasks requiring frequent client participation, such as FEMNIST, our method uses $55\%$ less communication than the 2nd best baseline (CFL). Similarly, on CITYSCAPES, CFLHKD outperforms HierFAVG by $66\%$, a critical advantage for bandwidth-constrained devices processing high resolution sensor data.   

    \item \textit{Faster convergence;} Bi-level aggregation coupled with inter-cluster knowledge sharing enables CFLHKD to resolve conflicting updates early and accelerate convergence without compromising accuracy. In dynamic environments such as CIFAR-10, CFLHKD achieves $10\times \; \& \; 24\times$ faster convergence than IFCA and FedProx, respectively.
 \end{itemize}

\begin{figure}[t]
  \centering
  \includegraphics[width=\columnwidth]{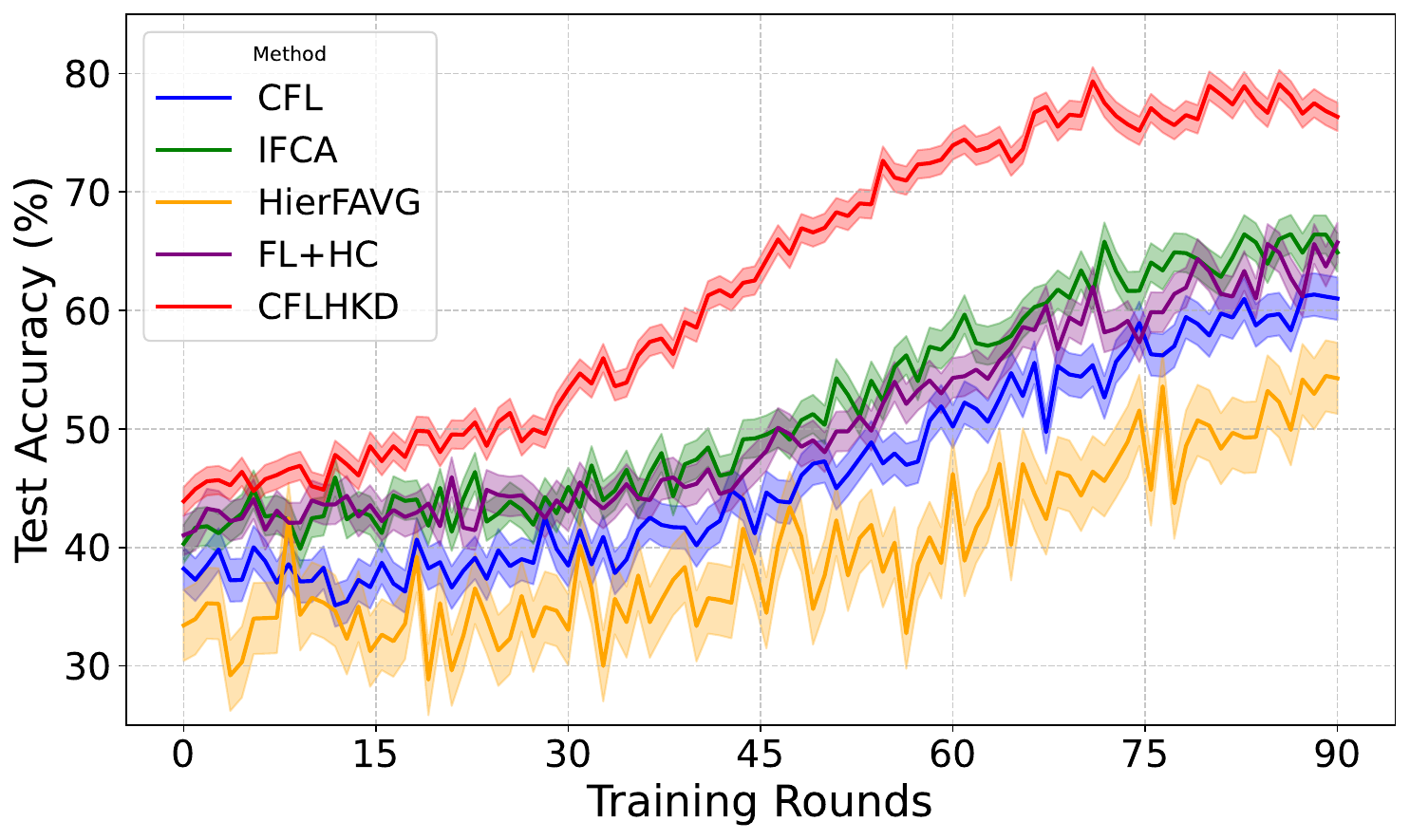}
  \caption{Cluster models accuracy over time.}
  \label{fig:cluster_models}
\end{figure}

Figure \ref{fig:cluster_models} depicts the evolution of test accuracy on HAM10000 dataset for individual clusters compared to a representative baseline of traditional, hierarchical, and clustered FL. In conventional CFL, clusters typically converge at disparate rates, with accuracy gaps exceeding $5-7\%$. On the other hand, CFLHKD not only demonstrates better accuracy but also converges faster. IFAC and FL+HC show competitive performance ($72.8$ \& $70.5$), with IFCA slightly outperforming FL+HC in later stages (after 60 rounds). While HierFAVG shows inconsistent performance, with accuracy dropping below $60\%$, indicating challenges in handling non-IID distributions without dynamic clustering and cross-cluster coordination. 

\subsubsection{Concept Drift} To evaluate robustness to distribution shift, we simulate label-based concept drift by changing the label distribution mid-training. For instance, on MNIST, clients initially train on digits 0-4, then abruptly shift to digits 5-9 at round 50.

Table \ref{tab:dynamic} demonstrates CFLHKD's remarkable robustness to concept drift, in two metrics including Accuracy Drop (Acc Drop); post drift performance drop and Recovery; training rounds required to regain pre-drift accuracy. For CIFAR-10, CFLHKD encounters a $4.8\%$ drop in accuracy, $62\% \; \& \; 74\%$ lower than IFCA and HierFAVG, respectively. In particular, it recovers after $3$ rounds, while FedAvg fails to recover after drift. Similarly, on HAM1000, CFLHKD experiences the smallest drop ($3.2\%, \; 69\%$ lower than IFCA) and fastest recovery (2 rounds) $3.5\times$ and $3\times$ faster than HierFAVG and IFCA. Dynamic cluster reassignment minimizes misaligned updates during drift while bi-level aggregation and inter-cluster knowledge sharing ensure quicker recovery after drift. On the other hand, conventional, hierarchical and CFL baselines struggle to handle and recover from concept drift (complete results in Table \ref{tab:dynamic_full}).

\subsubsection{Effectiveness of Inter-Cluster Knowledge Sharing} To evaluate the effectiveness of CFLHKD compared to conventional CFL, we analyze client model similarity matrices for CFLHKD and CFL on HAM10000 dataset. Both methods partition clients into four clusters (10 clients/cluster) using identical non-IID splits, and we compare their inter-cluster interactions at a fixed training round ($t = 100$).

\begin{figure}[!t]
  \centering
  \includegraphics[width=\columnwidth]{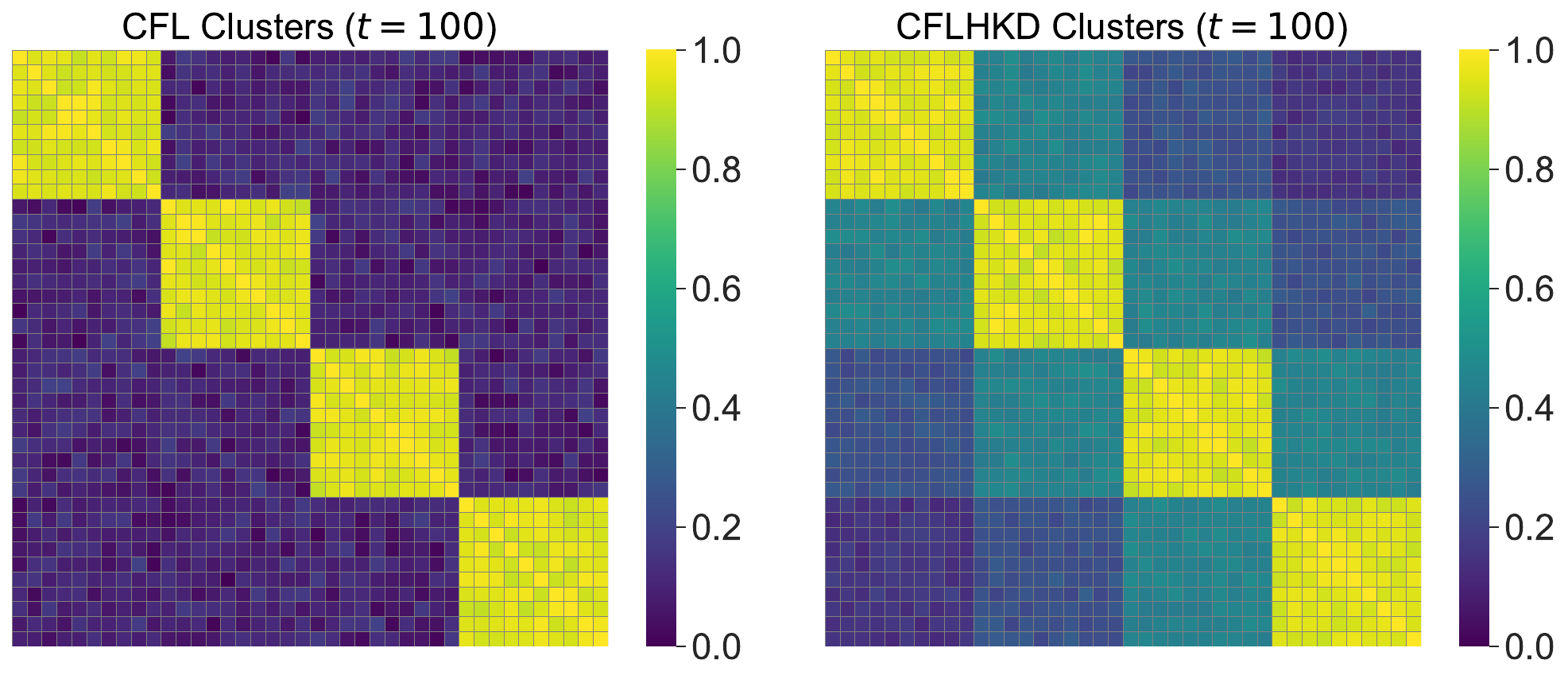}
  \caption{Cluster model divergence and inter-cluster knowledge sharing.}
  \label{fig:ickd}
\end{figure}

As shown in Figure \ref{fig:ickd}, CLF exhibits prominent block-diagonal patterns, indicating high intra-cluster similarity. However, the off-diagonal regions remain very dark, highlighting isolated learning with no meaningful exchange between clusters. In comparison, CFLHKD not only retains strong intra-cluster coherence but also maintains enhanced inter-cluster coordination. The off-diagonal regions show higher values (brighter than CFL), particularly among semantically related clusters; clusters 1-2 and 3-4 (Figure \ref{fig:ickd}). This controlled knowledge sharing ensures that related clusters benefit more while the rest benefit from shared high-level features without forcing them to over-align. CFLHKD mitigates fragmented learning by enabling structured cluster-aware knowledge exchange, resulting in $3.2\%$ accuracy gain over CFL (Table \ref{tab:overall}). Moreover, cross-cluster coordination reduces training time by $12\%$ compared to CFL.

\subsubsection{Scalability Study}
In this experiment, we evaluate the scalability of CFLHKD on two fronts: 1) cluster expansion (varying the number of clusters) and 2) client density (varying the number of clients per cluster).

\begin{figure}[t]
\begin{multicols}{2}
    
    \includegraphics[width=\linewidth]{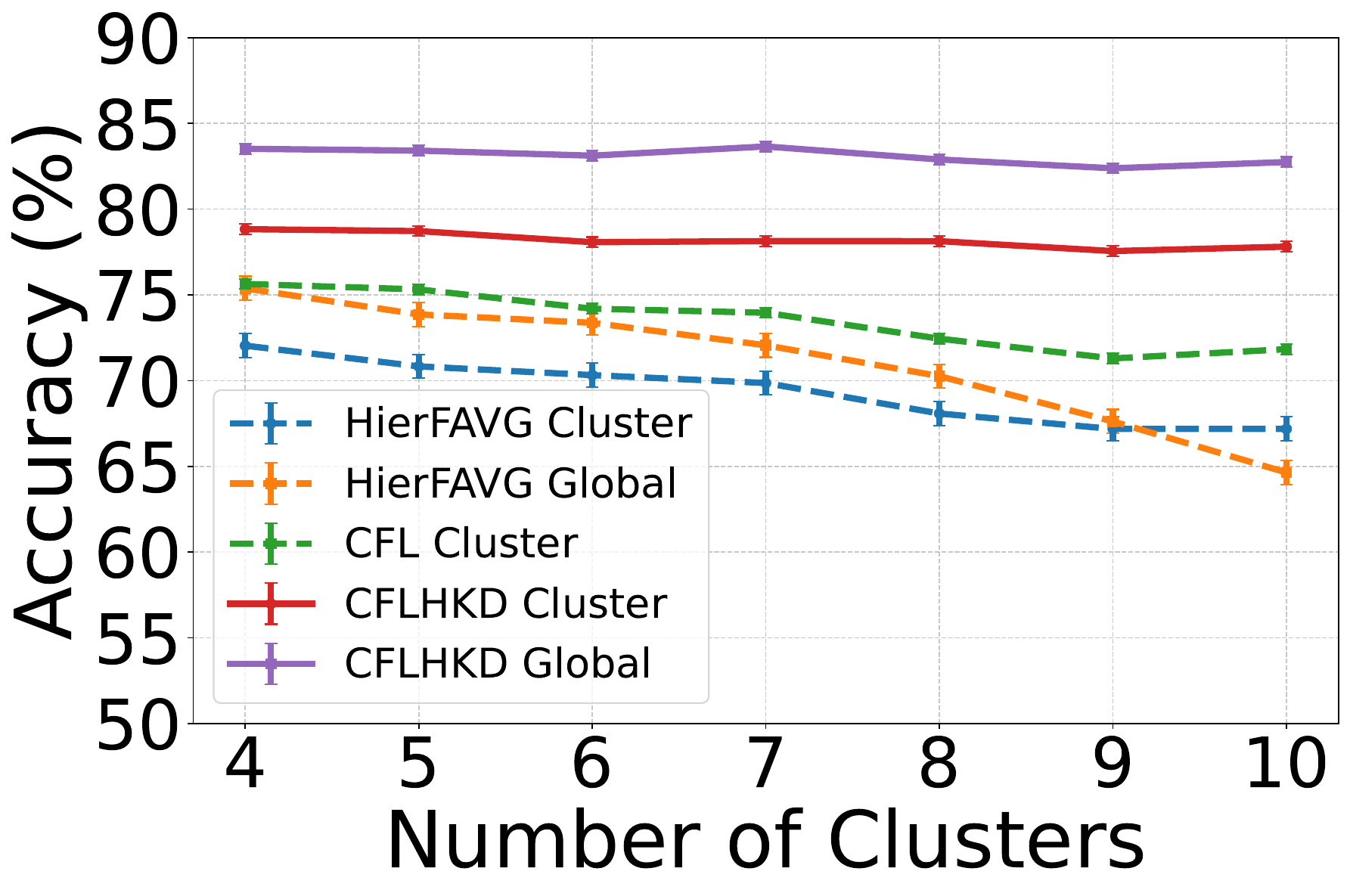}
    \caption{Number of clusters impact on performance}
    \label{fig:no_clusters}
    \par 
    \includegraphics[width=\linewidth]{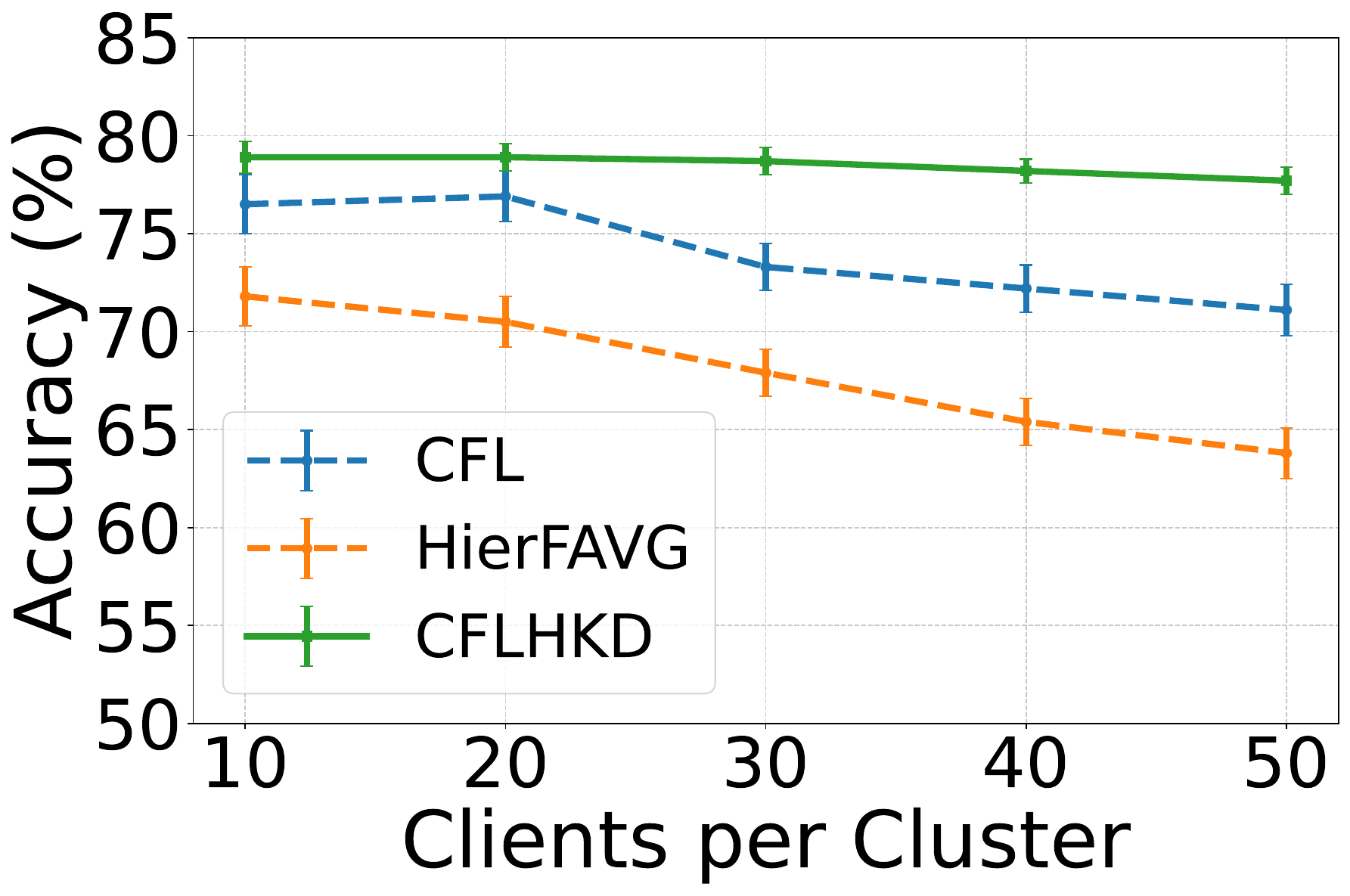}
    \caption{Density of cluster and its impact on performance.}
    \label{fig:ncluster_density}
    \par 
    \end{multicols}
\end{figure}

\textbf{Cluster expansion;} Figure \ref{fig:no_clusters} compares the impact of increasing the number of clusters (4 to 10) on the cluster and global model performance for CFLHKD, HierFAVG, and CFL (only cluster). As highlighted in Figure \ref{fig:no_clusters}, increasing the number of clusters exacerbates fragmentation as clusters become smaller, leading to a pronounced decline in the accuracies of global and cluster models. For instance, HierFAVG's cluster's accuracy drops from $75\%$ (4 clusters) to $67\%$ (10 clusters), a $12\%$ decline, while its global accuracy deteriorates even further ($14.7\%$) due to misaligned cluster models. In contrast, CFLHKD leverages bi-level aggregation and cross-cluster information exchange to mitigate fragmentation and exhibits a small decline in cluster and global accuracies with cluster expansion.

\textbf{Client density;} Figure \ref{fig:ncluster_density} evaluates the impact of increasing cluster size from $10$ to $50$ on cluster model performance. In CFL, higher cluster density lowers intra-cluster similarity while increasing difficulty to converge. Initially, CFL experiences a slight increase (for 20 clients/cluster) as it learns from an expanded pool; however, it faces a continual decline afterward (from $76\%$ to $70\%$). On the other hand, CFLHKD maintains relatively stable performance and only encounters a decline of $2.5\%$ even at $50$ clients per cluster.  

\begin{table}
    \centering
    \caption{Contribution of individual components of CFLHKD.}
    \label{tab:ablation}
    \renewcommand{\arraystretch}{1.2}
    \setlength{\tabcolsep}{8pt}
    \resizebox{\columnwidth}{!}{%
    \begin{tabular}{l|c|ccc|c|ccc}
        \toprule
        \multirow{2}{*}{Variant} & \multicolumn{4}{c|}{CIFAR-10} & \multicolumn{4}{c|}{CITYSCAPES} \\
        \cline{2-9}
        & Acc & Comm & Time & $\Delta$ & Acc & Comm & Time & $\Delta$  \\
        \midrule
         CFLHKD &  \textbf{82.3} & \textbf{89.6} & \textbf{6.3} & - & \textbf{74.5} & \textbf{89.6} & \textbf{14.4} & - \\
        \hline
        w/o Bi-level Aggregation &  77.8 & 174.7 & 8.5 & $\downarrow$4.5 &  69.8 & 174.7 & 19.2 & $\downarrow$4.7 \\
        w/o Global Fine-tuning & 79.1 & 123.6 & 7.0 & $\downarrow$3.2 & 72.1 & 123.6 & 16.0 & $\downarrow$2.4 \\
        w/o Dynamic Clustering & 80.5 & 89.6 & 6.6 & $\downarrow$1.8  & 73.2 & 89.6 & 15.0 & $\downarrow$1.3 \\
        \bottomrule
    \end{tabular}%
    }
\end{table}

\subsection{Ablation Study}
To quantify the impact of CFLHKD's core components, we systematically ablate its bi-level aggregation, global fine-tuning, and dynamic clustering mechanisms. Table \ref{tab:ablation} presents the performance of these variants against the original framework on CIFAR-10 and CITYSCAPES datasets.

\textit{Impact of bi-level aggregation;} CFLHKD leverages bi-level aggregation and MTKD to effectively fuse knowledge from cluster-specific models in a unified global model. Disabling bi-level aggregation results in the most severe performance degradation; $5.5\%$ and $6.3\%$ lower accuracy for CIFAR-10 and CITYSCAPES, respectively. Without client-edge coordination, clients transmit raw updates to the cloud, increasing both communication costs and convergence time, as clusters work in isolation. For instance, coordination without edge increases convergence time by $35\%$ and almost doubles the communications costs ($95\%$) on CIFAR-10.

\textit{Impact of global fine-tuning;} 
Disabling the global fine-tuning phase in which the unified global model is used to regularize and refine cluster models results in a significant drop ($3.8\% \; \& \; 3.2\%$ ) in accuracy for CIFAR-10 and CITYSCAPES. Distillation-free training increases inter-cluster variance as clusters fail to assimilate global insights. These results underscore that MTKD is crucial not only for improving accuracy via inter-cluster knowledge sharing but also for accelerating convergence and reducing communication costs.

\textit{Impact of dynamic clustering;} Finally, we study the impact of dynamic clustering on the overall system's performance. Although this component has little to no impact on communication costs, it is critical for maintaining a homogeneous set of clusters. Removing dynamic clustering results in a $2.2\%$ and $1.8\%$ drop in accuracy while increasing the convergence time by $4.7\% \; \& \; 4.2\%$ for CIFAR-10 and CITYSCAPES, respectively. 

\section{Conclusion}
\label{conclusion}
This paper presents CFLHKD, a hierarchical CFL framework that addresses the fragmented learning problem by bridging the gap between local and global learning. We systematically analyze the impact of the lack of a unified global model and cluster model divergence. We further propose bi-level hierarchical aggregation and MTKD to enable inter-cluster knowledge sharing. Extensive evaluations on diverse datasets show that CFLHKD outperforms traditional FL, HFL, and CFL baselines in accuracy and convergence and incurs $20-80\%$ less communication costs. These findings underscore the potential of CFLHKD as a scalable and adaptive solution for large-scale heterogeneous IoT environments characterized by dynamic evolving conditions.
  

\bibliographystyle{ACM-Reference-Format}
\bibliography{references}

\newpage
\appendix

\section{Appendix}
\subsection{Additional Experimental Settings}
\label{add_exp_settings}
We implement CFLHKD and baselines in Pytorch (2.0.1) with Flower for FL orchestration \cite{beutel2022flower}. Experiments run on NVIDIA T500, simulating edge-cloud hierarchies via multi-threaded processes.

\subsubsection{Training Setup} 
Each round involves $30\%$ of clients and each client trains for $5$ local epochs using SGD (momentum = 0.9, weight decay = $1 \times10^{-4}$) with a learning rate (LR) of $0.01$ decayed by $0.99$ every 20 rounds. The batch size is set to 32 for IC tasks and 16 for CityScapes to balance memory and gradient stability \cite{he2016deep}. Cluster-specific models are updated every 10 rounds while the global model is updated every 30 rounds. Models used; CNN(MNIST/FEMNIST), ResNet-18(CIFAR-10/HAM10000), and U-Net for CityScapes.

\subsubsection{Baselines:}
\begin{itemize}
    \item CFLHKD: regularization weights; $\lambda_1 = 0.1$ (clustering), $\lambda_2 = 0.005$ (model drift)
    \item FedProx \cite{li2020federated}: proximal term $\mu = 0.01$
    \item HierFAVG \cite{liu2020client}: edge aggregation every 5 rounds, cloud aggregation every 20 rounds.
    CFL\cite{ghosh2020efficient}/ICFL\cite{yan2023clustered}/IFCA\cite{ghosh2022efficient}: cluster update interval every 10 rounds, cosine similarity threshold for partitioning = 0.7.
    \item FL + HC \cite{briggs2020federated}: hierarchical cluster depth = 3, cluster reassignment frequency = 15 rounds.
\end{itemize}

All methods went independent grid search ($LR \in \{ 0.001, 0.01, 0.1\}$, $Local \; Epochs \in \{ 3, 5, 10\}$) on validation sets. CFLHKD's $\tau$ and aggregation interval were tuned separately.

\subsubsection{Dataset Configurations}
We experiment with two types of non-IID data configurations; static (standard non-IID) and dynamic to simulate concept drift.

\begin{itemize}
    \item Static non-IID (Label Skew); IC datasets (MNIST/FEMNIST/CIFAR-10) are partitioned via Dirichlet distribution ($\alpha = 0.5$) across 100 clients, following \cite{hsu2019measuring}. HAM10000 is split into 5 hospitals (20 clients/cluster) and CityScapes by cities (50 edge clusters), and each city contains 10-15 vehicles/clients per city \cite{fantauzzo2022feddrive}.

    \item Dynamic (Concept Drift); Clients within IC datasets switch labels at round 50 to simulate label shift \cite{liu2024fedca}. CityScapes clients experience gradual environmental changes (e.g., day-to-night or urban-to-rural) to simulate feature shift \label{fantauzzo2022feddrive}
\end{itemize}

\subsection{Case Study: CityScapes}
\begin{figure*}[t!]
\centering
\begin{minipage}{0.48\linewidth}
    \centering
    \includegraphics[width=\linewidth]{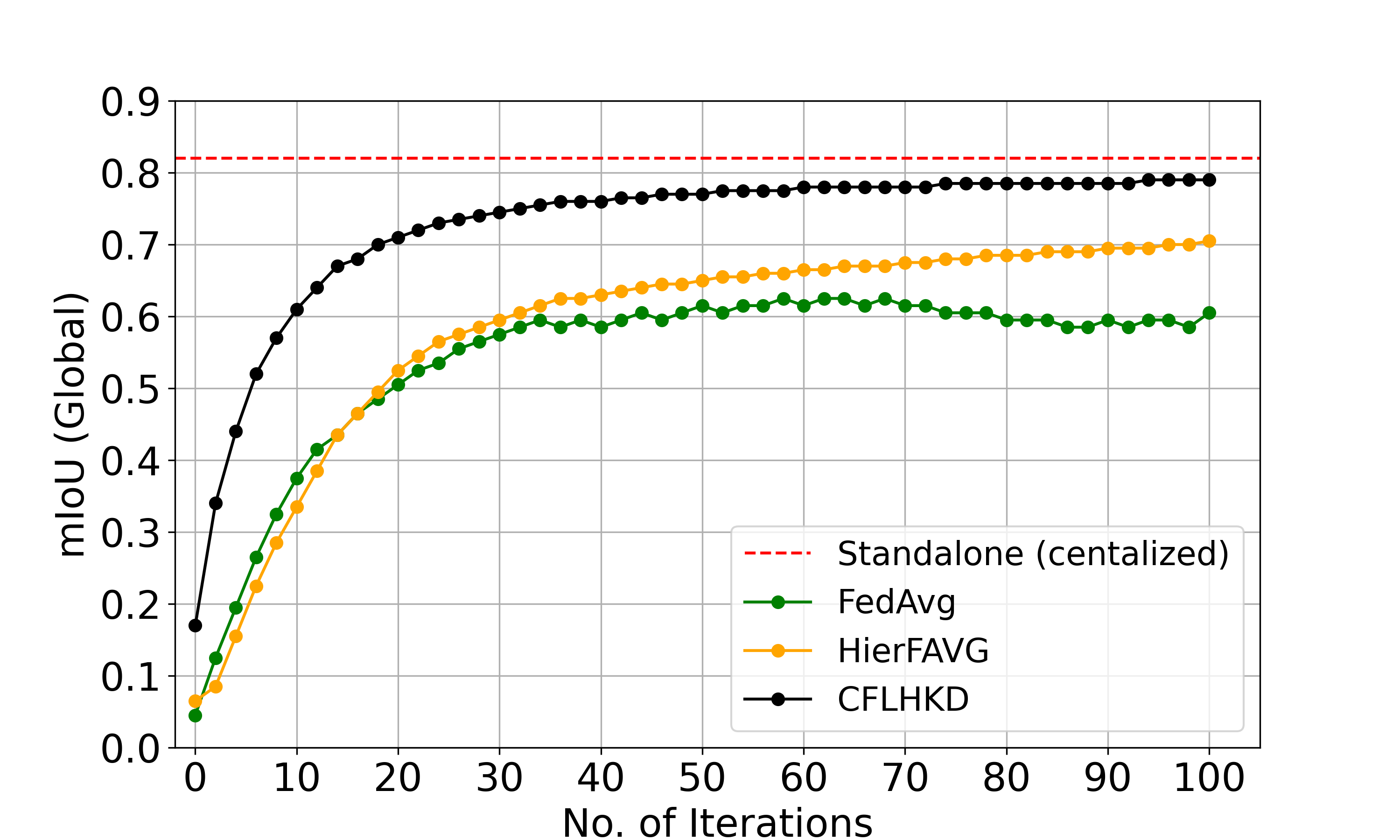}
    \caption{Comparison of global model's performance (standard non-IID).}
    \label{fig:g_miou}
\end{minipage}%
\hfill%
\begin{minipage}{0.48\linewidth}
    \centering
    \includegraphics[width=\linewidth]{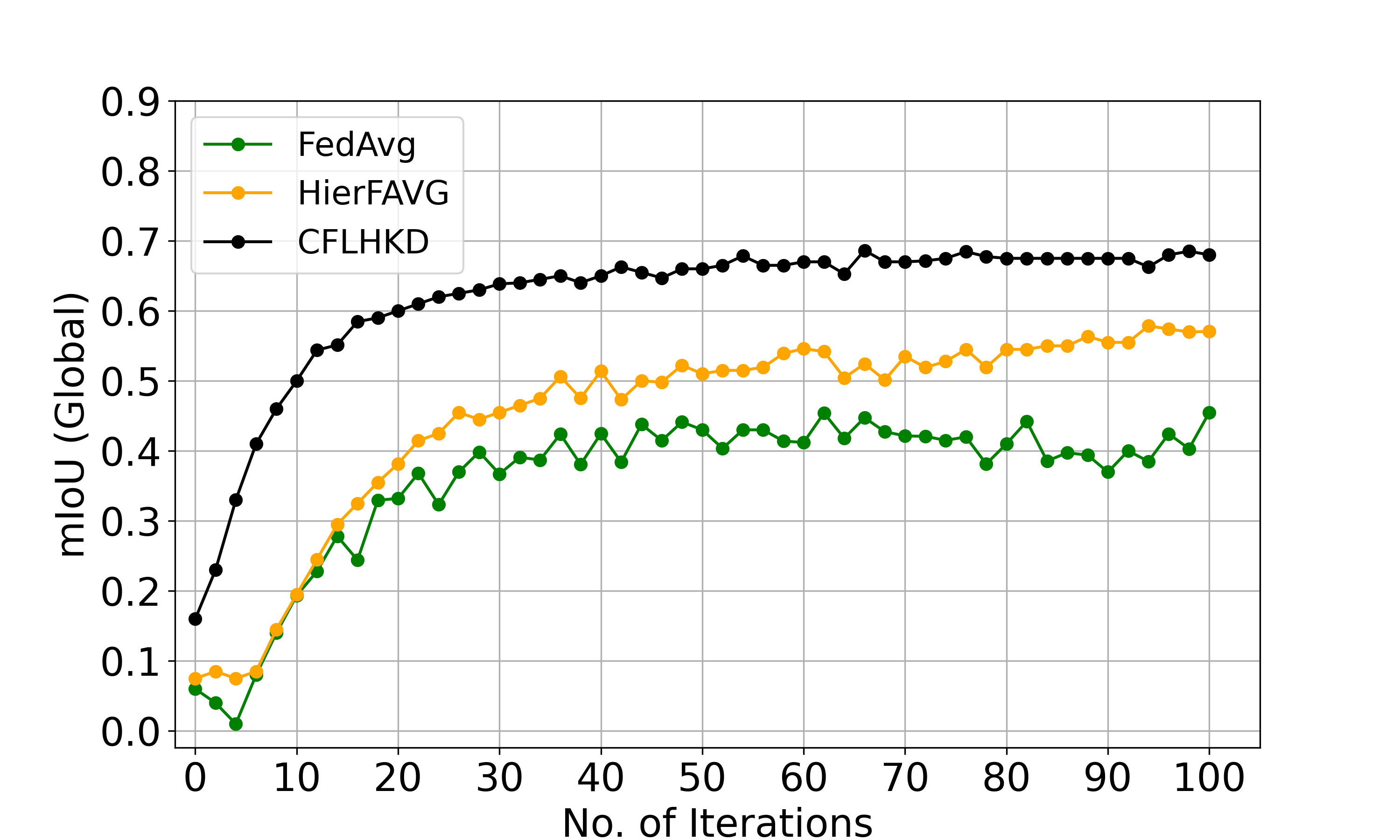}
    \caption{Comparison of global model's performance (concept drift).}
    \label{fig:loss}
\end{minipage}
\end{figure*}

\begin{figure*}[t!]
\centering
\begin{minipage}{0.48\linewidth}
    \centering
    \includegraphics[width=\linewidth]{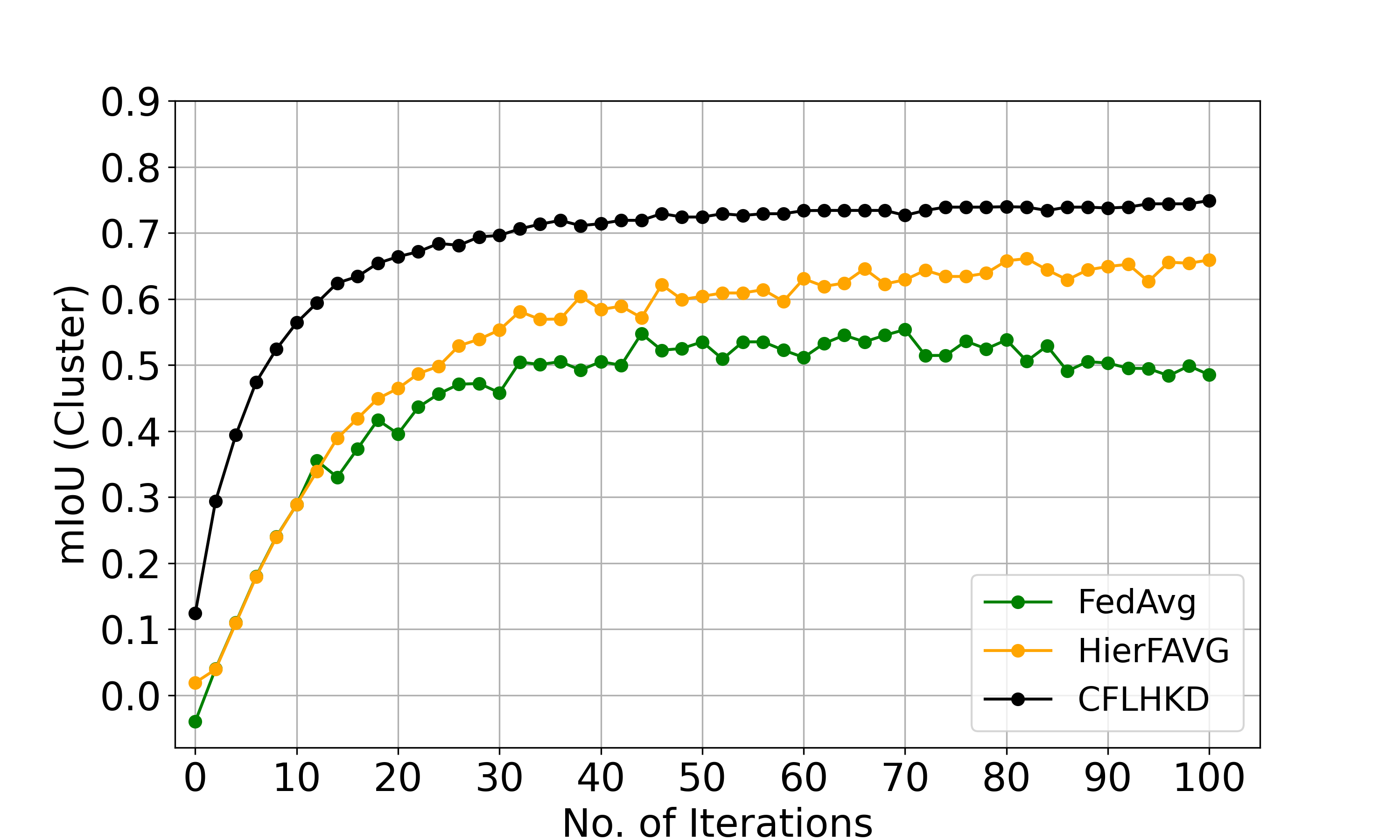}
    \caption{Cluster model performance (standard non-IID).}
    \label{fig:r_miou}
\end{minipage}%
\hfill%
\begin{minipage}{0.48\linewidth}
    \centering
    \includegraphics[width=\linewidth]{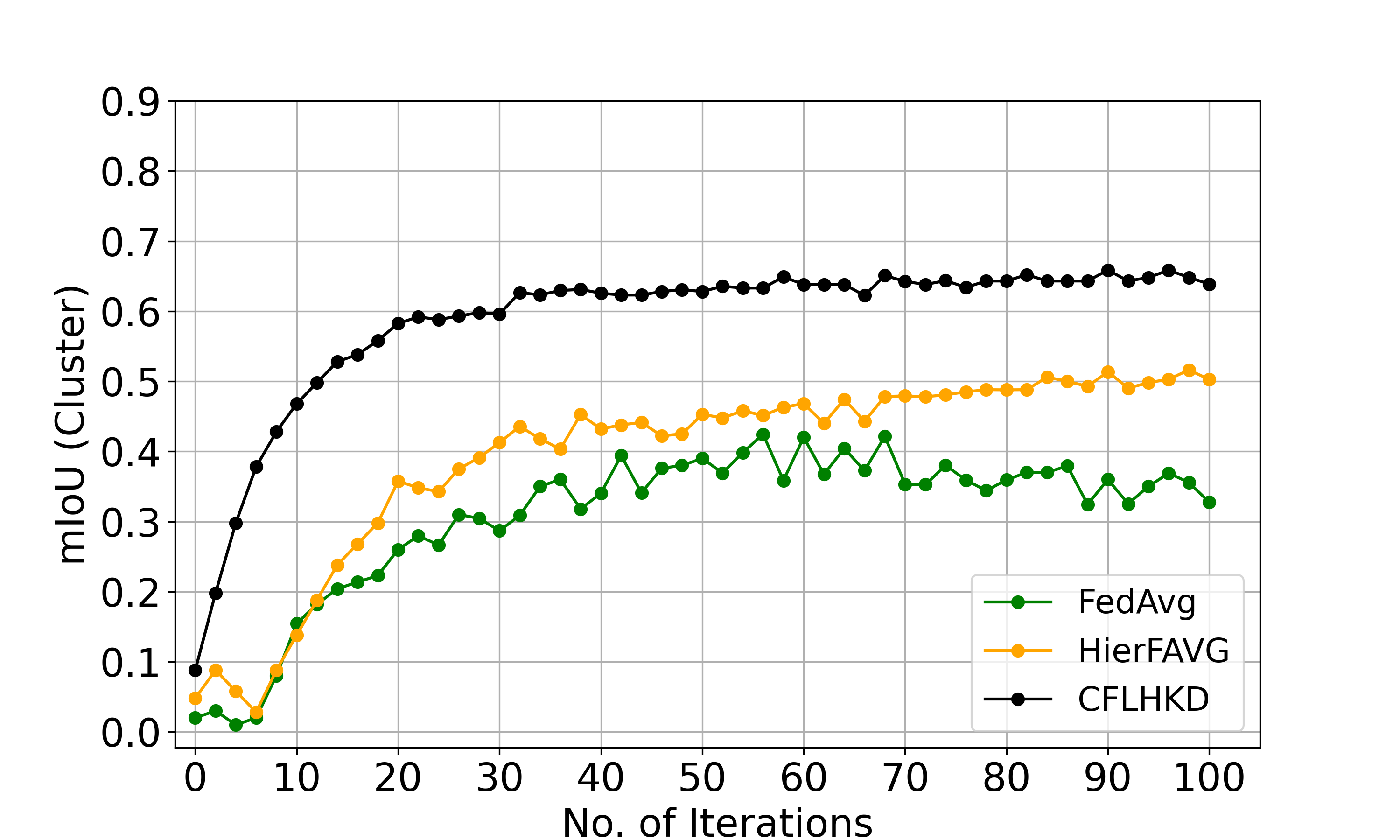}
    \caption{Cluster model performance (concept drift).}
    \label{fig:r_loss}
\end{minipage}
\end{figure*}

We conduct additional experiments on the CityScapes dataset to study the impact of evolving data distributions (concept drift) on the global and cluster-specific models performance and compare it with global and cluster models of HierFAVG baseline. These confirm the CFLHKD's ability to maintain effectiveness even in the face of concept drift as dynamic clustering and inter-cluster knowledge sharing help mitigate the deterioration of performance, as seen in the case of HierFAVG (see Figures \ref{fig:g_miou}, \ref{fig:loss}, \ref{fig:r_miou}, \ref{fig:r_loss}).

\subsection{Extended Ablation Study}
To validate the design choices and hyperparameter sensitivity of CFLHKD, we conduct a comprehensive ablation study on CIFAR-10 and CityScapes datasets. The results are averages over 3 seeds. As reported in Table \ref{tab:lambda_impact}, $\lambda = 0.1$ improves mIoU by $3.3\%$ without refinement, while a higher value ($0.5$) overregularizes and degrades local adaptation.

\subsubsection{Cluster Refinement Regularization ($\lambda$)}
$\lambda$ balances between cluster specialization ($\mathcal{L}(w_{e,k}, D_k)$) and global generalization ($||w_{e,k} - w_g||^2$) (Equation \ref{eq:ftl}).

\begin{table}[h]
    \centering
    \caption{CFLHKD under different values of $\lambda$.}\resizebox{\columnwidth}{!}{%
    \begin{tabular}{l|c}
        \toprule
        $\lambda$ & CityScapes (mIoU) \\
        \midrule
        0.0 (no refinement) & 72.1 \\
        0.1 & 74.5  \\
        0.5 & 72.7  \\
        \bottomrule
    \end{tabular}
    }
    \label{tab:lambda_impact}
\end{table}

\subsubsection{Affinity Trade-off ($\gamma$)}
$\gamma$ balances between data distribution (JSD) and model similarity for calculating affinity in cluster formation (Equation \ref{eq:affinity}). Hybrid affinity ($\gamma=0.5$ ) achieves the optimal cluster stability as reflected in the model's performance (see Table \ref{tab:gamma_impact}). 

\begin{table}[h]
    \centering
    \caption{CFLHKD under different values of $\gamma$.}\resizebox{\columnwidth}{!}{%
    \begin{tabular}{l|c}
        \toprule
        $\gamma$ & CityScapes (mIoU) \\
        \midrule
        0.0 (model) & 70.6 \\
        0.5 & 74.5  \\
        1.0 (data) & 69.3  \\
        \bottomrule
    \end{tabular}
    }
    \label{tab:gamma_impact}
\end{table}

\subsubsection{Clustering Threshold ($\phi$)}
$\phi$ is the threshold that determines client assignment to clusters based on affinity scores. $\phi = 0.7$ prevents over and under-clustering, maintaining high accuracy (Table \ref{tab:delta_impact}).

\begin{table}[h]
    \centering
    \caption{CFLHKD under different values of $\phi$.}\resizebox{\columnwidth}{!}{%
    \begin{tabular}{l|c|c}
        \toprule
        $\phi$ & CIFAR-10 (Acc) & Clusters (Initial/Final) \\
        \midrule
        0.3  & 87.1 & 4 $\rightarrow$ 8\\
        0.7 & 89.6 & 4 $\rightarrow$ 6 \\
        0.9  & 88.3 & 4 $\rightarrow$ 3 \\
        \bottomrule
    \end{tabular}
    }
    \label{tab:delta_impact}
\end{table}

\subsubsection{Complete Ablation Results}
We use the following formula for calculating the communication costs:

\begin{equation}
Comm = clients \times model \; size \times rounds \times aggregation \; frequency.    
\end{equation}

Table \ref{tab:extended_ablation} contains the complete results partially presented in Table \ref{tab:ablation}.

\begin{table*}[h]
    \centering
        \renewcommand{\arraystretch}{1} 
    \setlength{\tabcolsep}{2.5pt} 
    \caption{Dynamic Non-IID (Concept Drift)}
    \begin{tabular}{l|cc|cc|cc|cc|cc}
        \toprule
        \multirow{2}{*}{Method} & \multicolumn{2}{c|}{MNIST} & \multicolumn{2}{c|}{CIFAR-10} & \multicolumn{2}{c}{FEMNIST} & \multicolumn{2}{c}{HAM10000} & \multicolumn{2}{c}{CITYSCAPES}  \\
        \cline{2-11}
        & Acc Drop (\%) & Recovery & Acc Drop (\%) & Recovery & Acc Drop (\%) & Recovery  & Acc Drop (\%) & Recovery & Acc Drop (\%) & Recovery \\
        \midrule
        Standalone	& 14.2 &	- &	20.1 &	- &	18.5 &	- &	19.8 &	- &	12.3 &	- \\ \hline
        FedAvg    & 12.5 &	15 &	18.3	& -	& 16.7 & 20 & 15.8 &	20 & 9.5 & -  \\
        FedProx &	11.8 &	12 &	17.2 &	18 &	15.3 &	18 &	14.1 &	16 &	8.9 &	10 \\
        HierFAVG  & 10.5 &	10 &	15.6	& 15	& 14.1 &	15 & 13.2 & 14 & 7.8 & 8 \\
        FL+HC &	9.8 &	8 &	14.3 &	12	& 13.5 &	12	 & 12.4 &	12 &	6.7 &	6 \\
        CFL &	8.7 &	7 &	13.1 &	10 &	12.8 &	10 &	11.5 &	10 &	6.2 &	5 \\
        ICFL &	7.9 &	6 &	12.7 &	9	& 11.2 &	8 &	10.5 &	8 &	5.9 &	4 \\
        IFCA      & 8.2 &	6 &	12.7 &	8 & 10.9 & 7 &	10.5 &	7 & 6.4 & 4 \\
        CFLHKD   & \textbf{2.1} & \textbf{2} & \textbf{4.8} & \textbf{3} & \textbf{3.5} & \textbf{2}  & \textbf{3.2} & \textbf{2} & \textbf{1.7} & \textbf{2} \\
        \bottomrule
    \end{tabular}
    \label{tab:dynamic_full}
\end{table*}

\begin{table*}[h]
    \centering
    \caption{Ablation Study: extended results on additional datasets}
    \label{tab:extended_ablation}
    \renewcommand{\arraystretch}{1.2}
    \setlength{\tabcolsep}{8pt}
    \resizebox{\textwidth}{!}{%
    \begin{tabular}{l|c|ccc|c|ccc|c|ccc|c|ccc}
        \toprule
        \multirow{2}{*}{Variant} & \multicolumn{4}{c|}{MNIST} & \multicolumn{4}{c|}{CIFAR-10} & \multicolumn{4}{c|}{HAM10000} & \multicolumn{4}{c}{Cityscapes} \\
        \cline{2-17}
        & Acc & Comm & Time & $\Delta$ & Acc & Comm & Time & $\Delta$ & Acc & Comm & Time & $\Delta$ & mIoU & Comm & Time & $\Delta$ \\
        \midrule
         CFLHKD & \textbf{97.6} & \textbf{9.6} & \textbf{2.1} & - & \textbf{82.3} & \textbf{89.6} & \textbf{6.3} & - & \textbf{78.9} & \textbf{56.0} & \textbf{5.0} & - & \textbf{74.5} & \textbf{89.6} & \textbf{14.4} & - \\
        \hline
        w/o Bi-level Aggregation & 96.1 & 18.2 & 2.8 & $\downarrow$1.5 & 77.8 & 174.7 & 8.5 & $\downarrow$4.5 & 74.5 & 105.0 & 6.8 & $\downarrow$4.4 & 69.8 & 174.7 & 19.2 & $\downarrow$4.7 \\
        w/o Global Fine-tuning & 97.0 & 12.7 & 2.4 & $\downarrow$0.6 & 79.1 & 123.6 & 7.0 & $\downarrow$3.2 & 76.8 & 77.0 & 5.6 & $\downarrow$2.1 & 72.1 & 123.6 & 16.0 & $\downarrow$2.4 \\
        w/o Dynamic Clustering & 97.3 & 9.6 & 2.2 & $\downarrow$0.3 & 80.5 & 89.6 & 6.6 & $\downarrow$1.8 & 77.7 & 56.0 & 5.3 & $\downarrow$1.2 & 73.2 & 89.6 & 15.0 & $\downarrow$1.3 \\
        \bottomrule
    \end{tabular}%
    }
\end{table*}

\end{document}